\documentstyle[aps,multicol,preprint]{revtex}
\renewcommand{\vr}{\bbox{r}}
\newcommand{\vp}{\bbox{p}}
\newcommand{\vq}{\bbox{q}}
\newcommand{\vx}{\bbox{x}}
\newcommand{\vv}{\bbox{v}}
\newcommand{\vk}{\bbox{k}}
\newcommand{\cD}{{\cal D}}
\begin{document}
\draft
\title{On loss of quantum coherence in an interacting Fermi gas}
\author{Kasper Astrup Eriksen and Per Hedeg\aa rd}
\address{\O rsted Laboratory, Niels Bohr Institute for Astronomy, 
Physics and Geophysics\\
Universitetsparken 5, DK-2100 Copenhagen \O, Denmark}
\date{\today}
\maketitle
\begin{abstract}
We clarify the path integral calculation, recently suggested by Golubev 
and Zaikin \cite{GZ1,GZ2}, and show, contrary to their claim, that quasiparticles 
become fully coherent quantum particles in the $T \rightarrow 0$ limit. 
The important physical point is the inclusion of the recoil of the 
quasiparticle when interacting with fluctuations in the rest of the 
Fermi gas. 
\end{abstract}
\pacs{}
\section{Introduction}
Golubev and Zaikin (GZ) \cite{GZ1,GZ2} has recently made a rather dramatic proposal, that the
quasiparticles of an electron gas, will loose their quantum coherence in a 
finite --- not terribly long  --- time, even at temperatures going to 
the absolute zero, due to the interaction with the rest of the gas. 
Obviously such a fact would have far reaching consequences. One would 
have to rethink a number of fundamental properties of electron physics, 
in particular the theory of localization. And it is indeed to good to be 
true. In a very careful and pedagogical work Aleiner, Altshuler and 
Gershenson (AAS) \cite{AAS1,AAS2} have redone the theory of weak localization, this time taking 
electron-electron interactions into account. And they show convincingly, 
that electron-electron interaction does have an effect at $T=0$, but it is not 
to decohere the quasiparticles, but rather to give extra scattering off the 
static Friedel oscillation pattern of the screening cloud associated with the 
impurities. 

Still the GZ work is not without merits. It wants to formulate the many body 
transport theory problem using Feynman path integrals. This has several 
advantages. The physical picture emerging is much more readily understandable. 
The physics of weak localization itself was not widely known and accepted until 
the works of Altshuler et al.  \cite{Kmelnitskii}, Bergmann \cite{Bergmann} and Chakravarty and Schmid \cite{Chakravarty}, where a 
real space description in terms of electron paths was used. The generalization 
of such an approach to interacting many body systems is still not fully 
understood --- the GZ work is certainly an example of this. It therefore is of 
interest to find out, where exactly the GZ calculation goes wrong. The AAS 
calculation --- using standard methods of many body theory --- does show that it is 
wrong, but it does not really show how the procedure can be modified. The only 
hints that AAS offers is some vague statements about problems in semiclassical 
calculations when classical  orbits intersect. 

In this paper we want to show that the error in the GZ calculation is not so 
much related to the selfcrossing orbits relevant for weak localization, but 
rather stems from the fact that the recoil of the particle, who's path is being 
followed, in the real and virtual scattering processes with the rest of the gas, 
is not properly taken into account \cite{Imry1}. In section 2 we show how to set up a more 
realistic path integral calculation in the Caldeira-Leggett spirit also used by 
Golubev and Zaikin, and in the final section we show how scattering processes 
are properly dealt with, giving the result, that in the $T=0$ limit, the 
dephasing time $\tau_{\phi}$ for quasiparticles at the Fermi level will tend to 
$\infty$. 

\section{Path integrals of an interacting Fermi gas}
Our starting point is the same as that of Golubev and Zaikin. We want to consider an interacting electron gas, i.e. to consider a many body problem with the following Hamiltonian:
\begin{equation}
H = H_0 + H_{int},
\end{equation}
where 
\begin{equation}
H_0 = \int d\vr\; \psi^{\dagger}_\sigma(\vr)\left [-\frac{\nabla^2}{2m}-\mu+U(\vr)\right] \psi_\sigma(\vr),
\end{equation}
and
\begin{equation}
H_{int} = \frac{1}{2}\int d\vr\int d\vr'\; \psi^\dagger_\sigma(\vr)\psi^\dagger_{\sigma'}(\vr') v(\vr-\vr')\psi_{\sigma'}(\vr')\psi_\sigma(\vr)
\end{equation}
where $U(\vr)$ is an external potential which may or may not be random, $\mu$ is the chemical potential, and $v(\vr-\vr')$ is the interaction potential which we shall take to be the Coulomb potential --- others may be used, that will not affect the points we want to make in this paper. 

The goal is to calculate the many body density matrix, and from that get physical quantities like currents and densities. We shall follow GS and use a functional integral formulation, introduce a Hubbard-Stratonovic field $V_i(\vr,t)$ ($i$ being a Keldysh index), to decouple the electrons. 
The effective action for the fluctuating $V$-fields is calculated by GS to second order. Since the approximation of leaving out higher order terms is not being questioned, we shall do the same, and hence work with
\begin{eqnarray}
\lefteqn{iS[V_1,V_2] = i \int \frac{d\omega d^3 \vk}{(2\pi)^4}\left( V^-(-\omega, -\vk) \frac{k^2\epsilon(\omega, \vk)}{4\pi}V^+(\omega, \vk) \right .} \nonumber\\
&& +i \left . V^-(-\omega,-\vk)\frac{k^2\mbox{Im}\epsilon(\omega, \vk)}{4\pi}\coth\left(\frac{\omega}{2T}\right)V^-(\omega, \vk)\right ).
\end{eqnarray}
Here $V^+ = (V_1+V_2)/2$ and $V^-=V_1-V_2$. It is important to notice the inverse of the permitivity is a retarded function so for instance is the average $\langle V^-(-\vk,t')V^+(\vk,t'')\rangle$ zero for $t' > t''$. The physical density matrix $\rho (t)$ is obtained by averaging the density matrix in the presence of specific field history $\rho _V(t)$ over $V$
\begin{equation}
\delta \rho(t) = \frac{\int \cD V_1 \cD V_2 \delta \rho_V(t) e^{iS[V_1,V_2]}}{ \int \cD V_1 \cD V_2  e^{iS[V_1,V_2]}},
\end{equation}
$\rho _V(t)$ is given by
\begin{equation}
\rho _V(t, \vr, \vr ')\frac{\int d \overline{\psi} \int \cD \psi \cD \psi e ^{i S[\overline{\psi }, \psi , V;t]} 
\overline{\psi } _2 ( \vr ', t) \psi _1 (\vr, t)}{\int d \overline{\psi} \int \cD \psi \cD \psi e ^{i S[\overline{\psi }, \psi , V;t]} }
\end{equation}
with the effective electron action
\begin{equation}
S[\overline{\psi }, \psi , V; t ] = \int _{C_t} d t' \left( \int d \vr \left[i \overline{\psi }(t', \vr) \partial _{t'} \psi (t', vr) - e \overline{\psi }(t', \vr) \partial _{t'} \psi (t', vr) \right] - H_0\right)
\end{equation} 
The Keldysh contour $C _t$ runs from $- \infty$ up to time $t$ and then back to $- \infty$. By direct differentiation a differential equation for $\rho _V(t)$ can be derived and the linear response to an applied electrical potential $e V _x$ is then given by
\begin{equation}
i \frac{\partial \delta \rho _V(t)}{\partial t} = H_1 \delta \rho _V(t) - \delta \rho _V(t) H_2 - [e V_x, \rho _V]
\end{equation}  
The solution to this equation is (GZ Eq. (41))
\begin{equation}\label{densmat}
\delta\rho_V(t) = i \int_0^t dt'\,U_1(t,t')[eV_x,\rho_V(t')] U_2(t',t),
\end{equation}
where the time evolution operators $U_{1,2}$ are given by
\begin{equation}
U_{1,2}(t_1,t_2) = \mbox{T} \exp\left[-i\int_{t_1}^{t_2} dt' H_{1,2}(t')\right].
\end{equation}
and the effective Hamiltonians $H_{1,2}$ are functionals of the fluctuating Hubbard-Stratonovic fields and are given by 
\begin{eqnarray}
H_1 &=& H_0 - eV^+ - \frac{1}{2}(1-2\rho_0)eV^- \nonumber \\
H_2 &=& H_0 - eV^+ + \frac{1}{2} eV^- (1-2\rho_0),
\label{HamiltonianH1H2}
\end{eqnarray}
where $\rho_0$ is the equilibrium one particle density matrix
\begin{equation}
\rho_0 = \frac{1}{e^{\beta H_0}+1}.
\end{equation}
This whole theoretical setup is very close to the standard Caldeira-Leggett description of a single quantum particle in a dynamic environment, as also emphasized by Golubev and Zaikin in a more recent paper \cite{GZ3}, where they reply to some of the critizism raised. The environment is responsible for both a kind of ``dynamic Hartree'' contribution $-eV^+$ to the effective Hamiltion and a non-Hermitian contribution $- \frac{1}{2}(1-2\rho_0)eV^-$ (to $H_1$, and a similar to $H_2$). In contrast to the standard one-particle Caldeira-Leggett, the non-Hermitian contribution is very sensitive to the energy of the quasiparticle, through the factor $1-2\rho_0$, which essentially is a smoothed out sign function. This of course has to do with the Pauli principle playing an important role in the many body problem. 

At this point GS represent the one-particle time-evolution operators $U_1$ and $U_2$ as standard real space path integrals, i.e. the density matrix (\ref{densmat}) is an integral over pairs of paths. Quantum interferense effects, like the ones responsible for weak localization, only occur if the two paths in the pairs are allowed to differ substantially. Such pairs of different paths are suppressed due to the coupling to the $V$'s in the effective Hamiltonians. If the pairs differ, they will experience uncorrelated $V^\pm$ so, in order to study the strength of dephasing one only need to study one of the time-evolution operators $U_{1,2}$ and see how fast it decays. 

So we want to make a path integral representation of $U_1(t)$. As mentioned, the non-Hermitian part of the Hamiltonian is strongly energy dependent, so a path integral, where one has a lax attitude to the exact energy of the paths, as do GS when they in their paper make the approximation, that the relevant paths are semiclassical orbits (straight lines between impurity collisions) with constant energies. On the scale of the Fermi energy, $\epsilon_F$ this may be a good approximation, but a change of energy of order $k_BT$ in a electron-electron collision event may have a dramatic effect in the contribution from non-Hermitian term, which exactly varies on this much smaller energy scale. Mathematically the ordering of $\rho _0$ and $V ^{-}$ in (\ref{HamiltonianH1H2}) is very important, but this aspect is missed in a conventional path integral representation with classical orbits.  

To repair the calculation we on the one hand want to follow the eletrons in real space, since this is at the heart of e.g. weak localization, and on the other hand keep track of the energy and momentum to a precision much smaller than the Fermi energy. We therefore represent the one particle time evolution operator $U_1(t)$ as a path integral where we instead of using position eigenstates use wavepackets $|\vx,\vp\rangle$:
\begin{eqnarray}
\langle \vr | \vx,\vp\rangle &\propto& \exp\left(-\frac{(\vr-\vx)^2}{2 a^2}+ i \vp\cdot\vr\right), \nonumber \\ &&\mbox{real space wavefunction}\nonumber\\
\langle \vq | \vx,\vp\rangle &\propto& \exp\left(-\frac{a^2(\vp-\vq)^2}{2} + i (\vp-\vq)\cdot \vx\right),\nonumber \\ &&\mbox{momentum space wavefunction},
\end{eqnarray}
where the width $a$ will be chosen appropriately, e.g. 
\begin{equation}
a = \sqrt{\lambda_F l},
\end{equation} 
where $\lambda_F$ is the Fermi wavelength and $l$ is the eleastic mean free path. In this way we can locate the electron so well in space that it makes sense to talk about orbits that travel from one impurity to another. On the other hand the momentum (and hence the energy) is very well determined on the scale of the Fermi momentum and we can keep track of small changes in energy.

In this basis of wavepackets free propagation of an electron is given by
\begin{eqnarray}
\lefteqn{\langle \vp',\vx'| e^{-iH_0 t}|\vx, \vp\rangle }\nonumber \\
&=& \exp\left(-\frac{(\vx'-\vx-\vv t)^2}{4(a^2+\left(\frac{t}{2ma}\right)^2)}- \frac{(\vp-\vp')^2 a^2}{4}\right )\nonumber \\
&& \exp\left( -i\frac{\frac{1}{2}mv^2 t - \vk\cdot(\vx+\vx')/2 + \left(\frac{t}{2ma^2}\right)^2 \left(\frac{1}{2}m\left(\frac{\vx-\vx'}{t}\right)^2 t+\vp\cdot\vx-\vp'\cdot\vx'\right)}{1+\left(\frac{t}{2ma^2}\right)^2}\right ),
\end{eqnarray}
where 
\begin{equation}
\vv = \frac{\vp+\vp'}{2m}; \qquad \vk = \vp'-\vp.
\end{equation}
In the following we shall use the simple approximation
\begin{equation}\label{propagator}
\langle \vp',\vx'| e^{-iH_0 t}|\vx, \vp\rangle \approx \delta(\vx'-\vx - \vv t) \,\delta ( \vp-\vp')\;
 \exp\left(-i\epsilon_p t \right ),
\end{equation}
which of course is nothing but semiclassical propagation.

The matrixelements for the coupling to the rest of the electron gas is given by
\begin{eqnarray}\label{matrixelement1}
\lefteqn{\langle \vp',\vx'| eV^+(\vr,t)|\vx, \vp\rangle}  \nonumber \\
&&=
\int d\vk eV^+(\vk,t) \exp\left(-\frac{(\vx-\vx')^2}{4 a^2}- \frac{(\vp+\vk-\vp')^2 a^2}{4} - i (\vp+\vk-\vp')\cdot(\vx+\vx')/2\right )
\nonumber \\
&&\approx \delta(\vx-\vx') \int d\vk\; \delta(\vp+\vk-\vp')\; V^+(\vk,t)
\end{eqnarray}
and
\begin{eqnarray}\label{matrixelement2}
\lefteqn{\langle \vp',\vx'|\frac{1}{2}(1-2\rho_0)eV^-(\vr,t)|\vx, \vp\rangle }\nonumber \\
&&= \int d\vq \int d\vk \exp\left( -\frac{(\vp'-\vp-\vk)^2 a^2}{4} - \frac{(\vx -\vx')^2}{4a^2} - i (\vp'-\vp-\vk)\cdot \frac{\vx+\vx'}{2} - a^2 \left(q - i \frac{(x' - x)^2}{2 a^2} \right)^2 \right ) \nonumber \\
&&\qquad\qquad \frac{1}{2}(1-2\rho_0(\epsilon_{\frac{p' + p + k}{2}+q}))eV^-(\vk,t)\nonumber \\
&&\approx \delta(\vx-\vx')\;\int d\vk \;\delta(\vp'-\vp-\vk)\;\frac{1}{2}(1-2\rho_0(\epsilon_{p+k}))eV^-(\vk,t).
\end{eqnarray}

We now carry out the standard path integral construction of slicing the time into $N$ infinitesimal pieces $\delta t$, and insert an (over-)complete set of wavepackets between the $U_1(\delta t)$ factors:
\begin{eqnarray}\label{pathintegral}
\lefteqn{I = \int \cD V \langle \vp_f, \vx_f | U_1(t) | \vx_i, \vp_i \rangle e^{iS[V_1,V_2]}}\nonumber \\
&=&  \int\cD V \prod_n \int d\vx_n \int d\vp_n \langle \vp_{n+1},\vx_{n+1} | e^{-iH(t_n)\delta t}|\vx_n,\vp_n\rangle e^{iS[V_1,V_2]}.
\end{eqnarray}
In this expression we want to integrate out both momenta and the fluctuating $V$ fields, leaving us with a real space path integral. To obtain the effective action for that integral, we use a variation of Pauli's standard $\lambda$ trick. First the interaction ($H' = H - H_0$) is multiplied by a factor $\lambda$. Then we consider a functional $g(\lambda,\vx)$ of the real space orbits:
\begin{equation}
g(\lambda,\vx) = \int \cD V \prod_n \int d\vp_n \langle \vp_{n+1},\vx_{n+1}| e^{-iH^ \lambda(t_n)\delta t}|\vx_n,\vp_n\rangle e^{iS[V_1,V_2]}.
\end{equation}
This functional satisfy a simple differential equation in $\lambda$:
\begin{equation}\label{g-ligning}
\frac{\partial g(\lambda,\vx)}{\partial\lambda} = -i \int_0^t dt' \frac{\langle\langle U^ \lambda(t,t',\vx)H'(t',\vx)U^ \lambda(t',0,\vx)\rangle\rangle}{\langle\langle U^ \lambda(t,0,\vx)\rangle\rangle} g(\lambda,\vx),
\end{equation}
where we have taken the $N\rightarrow\infty$ limit, introduced the notation
\begin{equation}
\langle\langle \cdot \rangle \rangle = \int \cD V\int d\vp\; \cdot e^{iS[V_1,V_2]},
\end{equation}
and the time evolution function
\begin{equation}
U^ \lambda(t_2,t_1,\vx) = \prod_{n=n_1}^{n_2} \langle \vp_{n+1},\vx_{n+1}|e^{-iH^ \lambda(t_n)\delta t} |\vx_n,\vp_n\rangle, \quad n_i = t_i/t\, N.
\end{equation}
The equation (\ref{g-ligning}) is easily solved, and we end up with the following expression for the path integral (\ref{pathintegral})
\begin{equation}
I = \int \cD x\, g(0,x)\,e^{i S_{eff}(x)},
\end{equation}
where the effective action is
\begin{equation}
S_{eff}(\vx) = - \int_0^1 d\lambda \int_0^t dt' \frac{\langle\langle U^ \lambda(t,t',\vx)H'(t',\vx)U^ \lambda(t',0,\vx)\rangle\rangle}{\langle\langle U^ \lambda(t,0,\vx)\rangle\rangle}.
\end{equation}
Like GZ we are going to evaluate this action to second order in the interaction, i.e. expanding the time evolution operators to only first order in $\lambda$, since $S_{eff}$ contains an explicit interaction term. The final result becomes
\begin{equation}
S_{eff}(\vx) = i \int_0^t dt' \int_0^{t'} dt'' \frac{\langle\langle U^0(t,t',\vx)H'(t')U^0(t',t'',\vx)H'(t'')U^0(t'',0,\vx)\rangle\rangle}{\langle\langle U^0(t,0,\vx)\rangle\rangle }.
\end{equation}
This action is a functional of the real space path being considered. For simplicity we shall evaluate it for straight line path with a constant momentum $\vp$ very close to the Fermi momentum. 
Using the semiclassical propagator (\ref{propagator}) and the matrix elements (\ref{matrixelement1}) and (\ref{matrixelement2}) this action is readily evaluated to 
\begin{eqnarray}\label{seff1}
S_{eff}(\vx) &=& \int_0^t dt'\int_0^{t'} dt'' \int d\vk \left (\langle V^+(\vk,t')V^+(-\vk,t'')\rangle + \frac{1}{2}(1-2\rho_0(\epsilon_{p-k}))\langle V^+(\vk,t')V^-(-\vk,t'')\rangle \right )\nonumber \\ 
&&e^{-i(\epsilon_{p-k}-\epsilon_p)(t'-t'')} \nonumber\\
&=& \int d\vk \int \frac{d\omega}{2\pi} \left (\langle V^+(\vk, \omega)V^+(-\vk, -\omega)\rangle + \frac{1}{2}(1-2\rho_0(\epsilon_{p-k}))\langle V^+(\vk,\omega)V^-(-\vk,-\omega)\rangle \right ) \nonumber \\ &&f(-\omega +\vv \cdot \vk, t).
\end{eqnarray}
Here we have used that $\langle V^-(-\vk,t')V^+(\vk,t'')\rangle$ is an advanced function and hence zero in our context. 
The function $f(\Omega,t)$ carries the time dependence and is given by
\begin{equation}
f(\Omega,t) = \int_0^t dt'\int_0^{t'} dt'' e^{i\Omega(t'-t'')} \nonumber \\
= it\left(\frac{1}{\Omega} - \frac{\sin\Omega t}{\Omega^2 t}\right ) + t \frac{1-\cos \Omega t}{\Omega^2 t}.
\end{equation}

At this point we pause and make a detailed comparison to the GZ calculation. The last term of (\ref{seff1}) is equivalent to the first term in GZ eq. (54), taking into account, that we only calculate the contribution to the action from one path. If the GZ function $R$ is written out, one obtains in the GZ version
(using $\vr(t_1)-\vr(t_2) = \vv (t_1-t_2)$)
\begin{equation}
\int d\vk \int \frac{d\omega}{2\pi} \frac{4\pi}{k^2\epsilon(\omega,\vk)} \left( 1-2n(\vp)\right ) f(\omega+\vv\cdot\vk,t),
\end{equation}
and in our version, with $\langle V^+(-\vk,-\omega)V^-(\vk,\omega)\rangle$ written out
\begin{equation}
\int d\vk \int \frac{d\omega}{2\pi} \frac{4\pi}{k^2\epsilon(\omega,\vk)} \left( 1-2n(\vp+\vk)\right )f(\omega+\vv\cdot\vk,t),
\end{equation}
where we now use the GZ notation $n(\vp) = \rho_0(\epsilon_p)$. The difference is, that in our formula the Pauli factor depends on the recoiled momentum $\vp+\vk$ whereas in GZ this factor is depending on $\vp$! This we believe is precisely where the GZ calculation goes wrong.

We shall now continue with our calculation, and show that it gives the expected result, i.e. no dephasing for electrons at the Fermi level at $T=0$.

First, for $f(\Omega,t)$ in the long time limit we have
\begin{eqnarray}
f(\Omega,t) &=& \int_0^t dt'\int_0^{t'} dt'' e^{i\Omega(t'-t'')} \nonumber \\
&=& it\left(\frac{1}{\Omega} - \frac{\sin\Omega t}{\Omega^2 t}\right ) + t \frac{1-\cos \Omega t}{\Omega^2 t} \nonumber \\
&\approx& it\left(\frac{P}{\Omega} -i \pi\delta(\Omega)\right )\qquad\mbox{large $t$} \nonumber \\
&=& \frac{it}{\Omega + i\eta}\qquad\mbox{$\eta$ a positive infinitesimal}.
\end{eqnarray}
Since $\langle V^+(-\vk,-\omega)V^-(\vk,\omega)\rangle$ is the Fourier transform of a retarded function, it is analytical in the lower complex $\omega$ plane, and the $\omega$ integral in (\ref{seff1}) is performed by closing the contour in the lower half plane and picking up a contribution from the pole at $\omega= -\vk\cdot\vv - i\eta$. 

Since our main concern is with the loss of coherence, we shall only evaluate the imaginary part of $S_{eff}$. After some simple manipulations we get our final and main result in the long time limit
\begin{equation}
\mbox{Im}S_{eff}(\vx) = t \int d\vk \frac{4\pi}{k^2} \frac{\mbox{Im}\epsilon(\vk\cdot\vv,\vk)}{|\epsilon(\vk\cdot\vv,\vk)|^2}\left ( \coth\left(\frac{\vk\cdot\vv}{2k_BT}\right)-
\tanh\left(\frac{\vk\cdot\vv - \epsilon_p}{2k_B T}\right )\right )
\end{equation}
In the zero temperature limit, which is our main concern this reduces to
\begin{equation}
\mbox{Im}S_{eff}(\vx) = t \int d\vk \frac{8\pi}{k^2} \frac{\mbox{Im}\epsilon(\vk\cdot\vv,\vk)}{|\epsilon(\vk\cdot\vv,\vk)|^2}\Theta(\epsilon_p^2-(\vk\cdot\vv)^2)\mbox{sign}(\vv\cdot\vk).
\end{equation}
This is by no means surprising. We actually have recovered the standard result from standard many body theory. The coefficient to $t$ is nothing but the imaginary part of the self energy of a particle with momentum $\vp$. This does go to zero when the momentum approaches the Fermi momentum, resulting in infintely long lifetimes --- and coherence times --- for a quasiparticle at the Fermi surface!

\section{Conclusion} 
The claim by Golubev and Zaikin, that quasiparticles at the Fermi level will decohere due to the interaction with the zero-point fluctuations of the modes of the rest of the Fermi gas has been shown to be due to an error when dealing with the real space path integral representation of quasiparticle propagation. If one does not allow for tiny changes in the momentum and energy --- not significant for the overall path, who's shape is dominated by the fast velocity, $v_F$ --- then the erroneous result occur. The cure is to use wavepackets, with reasonably well defined space and momentum coordinates. The space part of the resulting path integral takes care of the overall propagation of the particle, whereas the momentum coordinate is important when calculating the interaction with low energy modes of the rest of the system. This interaction is very much dominated by the Pauli exclusion principle, which is extremely sensitive to minute changes (on the scale of $\epsilon_F$) in the energy. The latter effects are of course much more precisely described in standard many body theory, which usually uses a basis of momentum eigenstates. 

From our point of view the story is now over. We have shown that quasiparticles can indeed stay coherent and e.g. give an interference pattern in a double slit experiment. In such an experiment the paths going through one or the other slit are very different, making our considerations with only one path relevant. GZ are studying weak localization, where the interfering paths are time reversed paths, which actually visit the same spatial points (but at different times), giving an unlikely loophole for such paths to decohere. In an appendix we show that also these time reversed paths interfere!

\section{Acknowldgements} We thank Vinay Ambegaokar for a number of useful comments concerning our manuscript. KAE is supported by the Danish Natural Science Research
Council through Ole R\o mer Grant No.\ 9600548.

\appendix %
\section*{}

For the forward in time propagator $U_1$ we found the effective action due to the interaction (\ref{seff1})
\begin{eqnarray}
S_{eff}(\vx) &=& \int_0^t dt'\int_0^{t'} dt'' \int d\vk \left (\langle V^+(\vk,t')V^+(-\vk,t'')\rangle + \frac{1}{2}(1-2\rho_0(\epsilon_{p-k}))\langle V^+(\vk,t')V^-(-\vk,t'')\rangle \right )\nonumber \\ 
&&e^{-i(\epsilon_{p-k}-\epsilon_p)(t'-t'')} \nonumber\\
&=& \int d\vk \int \frac{d\omega}{2\pi} \left (\langle V^+(\vk, \omega)V^+(-\vk, -\omega)\rangle + \frac{1}{2}(1-2\rho_0(\epsilon_{p-k}))\langle V^+(\vk,\omega)V^-(-\vk,-\omega)\rangle \right ) \nonumber \\ &&f(-\omega +\vv \cdot \vk, t).
\end{eqnarray}
Likewise we find for $U_2$ the effective action
\begin{eqnarray}
S_{eff}(\vx) &=& \int_0^t dt'\int_0^{t'} dt'' \int d\vk \left (\langle V^+(\vk,t')V^+(-\vk,t'')\rangle - \frac{1}{2}(1-2\rho_0(\epsilon_{p+k}))\langle V^+(\vk,t')V^-(-\vk,t'')\rangle \right )\nonumber \\ 
&&e^{i(\epsilon_{p+k}-\epsilon_p)(t'-t'')} \nonumber\\
&=& \int d\vk \int \frac{d\omega}{2\pi} \left (\langle V^+(\vk, \omega)V^+(-\vk, -\omega)\rangle - \frac{1}{2}(1-2\rho_0(\epsilon_{p+k}))\langle V^+(\vk,\omega)V^-(-\vk,-\omega)\rangle \right ) \nonumber \\ &&f(-\omega +\vv \cdot \vk, t).
\end{eqnarray}
The pairs of paths that contribute to the weak localization effect are the ones where forward $1$ and backward in time $2$ paths are each others time reversed partners. In this case it is also possible to get cross contributions where one $V$ is on the forward path while the other is on the backward in time path. These contributions correspond in GZ to the two last terms in (54) and (55) and after focusing on the weak localization (68) and (69). Like GZ we find that the cross contributions disappear after averaging over the diffusive paths. The remaining term in GZ (69) originates from the $V^+$ $V^+$ average in the above actions and is identical to what GZ find in (71). In GZ the first term in (68) vanishes. We find in analogy  with the above single line calculation an extra contribution proportional to $\tanh\left(\frac{\vk\cdot\vv - \epsilon_p}{2k_B T}\right )$. To see this explicitly let us consider a particular straight line segment of the path and let us consider the case where the interaction also takes place within this straight line segment. If the velocity along the forward path $1$ is $\vv$ is the velocity along the time reversed path $2$ $- \vv$. The combined contribution from the second term in the effective actions is
\begin{equation}
\int d\vk \int \frac{d\omega}{2\pi} (\rho_0(\epsilon_{\vp + \vk})-\rho_0(\epsilon_{\vp-\vk}))\langle V^+(\vk,\omega)V^-(-\vk,-\omega)\rangle  f(-\omega +\vv \cdot \vk, t).
\end{equation}

We have used $\langle V^+(\vk,\omega)V^-(-\vk,-\omega)\rangle $ is even in $\vk$ and $\epsilon_{-p-k} = \epsilon_{p+k}$ 

Compared to GZ (68) we see that the only difference is that instead of a factor $(\rho_0(\epsilon_{p})-\rho_0(\epsilon_{p})) = 0$ we get 
\begin{equation}
(\rho_0(\epsilon_{p +k})-\rho_0(\epsilon_{p-k})) = - \frac{\tanh\left(\frac{\vk\cdot\vv - \epsilon_p}{2k_B T}\right ) - \tanh\left(\frac{-\vk\cdot\vv - \epsilon_p}{2k_B T}\right )}{2},
\end{equation}
 so in GZ the $\coth\left(\frac{\omega}{2k_B T}\right)$ should always be replaced with 
\begin{equation}
\coth\left(\frac{\omega}{2k_BT}\right)-
\tanh\left(\frac{\omega - \epsilon_p}{2k_B T}\right ),
\end{equation}
which vanshes in the $T\rightarrow 0$ and $\epsilon_p\rightarrow 0$ limit.

\end{document}